# Top-Down Transaction-Level Design with TL-Verilog

Steven Hoover
Redwood EDA
Shrewsbury, MA, USA
steve.hoover@redwoodeda.com

Ahmed Salman
Alexandria, Egypt
e.ahmedsalman@gmail.com

*Abstract*—Transaction-Level Verilog (TL-Verilog) is an emerging extension to SystemVerilog that supports a new design methodology, called *transaction-level design*. A *transaction*, in this methodology, is an entity that moves through structures like pipelines, arbiters, and queues. A transaction might be a machine instruction, a flit of a packet, or a memory read/write. *Transaction logic*, like packet header decode or instruction execution, that operates on the transaction can be placed anywhere along the transaction's flow. Tools produce the logic to carry signals through their flows to stitch the transaction logic.

We implemented a small library of TL-Verilog flow components, and we illustrate the use of these components in a top-down design methodology. We construct a hypothetical microarchitecture simply by instantiating components. Within the flows created by these components, we add combinational transaction logic, enabling verification activities and performance evaluation to begin. We then refine the model by positioning the transaction logic within its flow to produce a high-quality register-transfer-level (RTL) implementation.

*Keywords—electronic design automation; hardware description language; register-transfer logic; digital logic; integrated circuit; high-level modeling; transaction-level modeling; top-down design methodology; aspect-oriented programming; productivity; design library*

## I. INTRODUCTION

While circuit complexity continues to scale exponentially, it is not practical for the size of design teams nor the duration of projects to scale accordingly. This necessitates continual adoption of higher-level design practices. Register-transfer-level (RTL) modeling, however, has remained the norm for over 30 years and 1/25,000 transistor scaling. Design teams continue to grow, projects are limited in scope, and project durations lengthen. The industry is at its breaking point. While adoption of high-level modeling methodologies is necessary, it has been slow by the facts that high-level modeling languages are failing to offer sufficient simplification to justify the loss of direct control over details.

Recently, a *timing-abstract* design methodology has been demonstrated which preserves RT-level control while enabling designs to remain abstract [1]. Timing-abstraction is provided in Transaction-Level Verilog (TL-Verilog) by a pipeline construct. All logic belongs to a pipeline and exists within a pipeline stage. The stage of a logic expression is purely a physical implementation attribute. It has no impact on behavior, so timing details can be refined safely without impacting behavior. While the resulting expression of the design contains identical detail, it is simpler, smaller, and both easier and safer to manipulate.

*Transaction-level design* builds upon timing-abstract design. It extends the notion of transactions flowing through pipelines to transactions flowing through arbitrary "flow elements," such as FIFOs, arbiters, and queues.

This paper presents an early view of the use of transaction-level design using TL-Verilog in a top-down design methodology. As a showcase, we developed a rudimentary library of microarchitectural components [5], and we used them to construct and refine a hypothetical microarchitecture. Though we believe this showcase to be reflective of real-world use, we leave real-world examples to future research. We study a top-down approach not to imply that transaction-level design *is* a top-down methodology, but rather to demonstrate that a top-down approach is, indeed, possible. Top-down design has long been explored, but it is, today, generally understood to be infeasible. We show that it is not.

The remainder of the paper is organized as follows. Section II describes transaction-level design methodology in TL-Verilog. Section III describes a property of TL-Verilog termed "lexical reentrace". Section IV describes how transaction-level design with lexical reentrance enables the development of library components that are infeasible in RTL and the use of these components in a top-down design flow. Section V quantifies the results, followed by conclusions in Section VI.

## II. TRANSACTION-LEVEL DESIGN

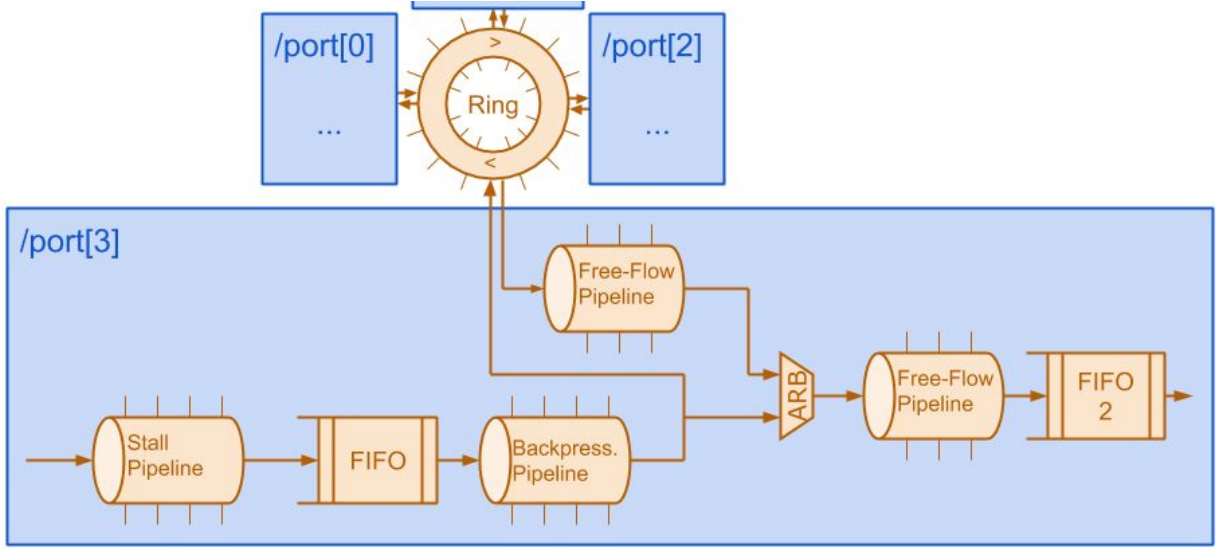

Fig. 1. Example transaction flow

Our showcase design is depicted in Fig. 1. It implements a *transaction flow* from various *flow components* in our library. This design connects four equivalent logic blocks (ports) on a ring. Each cycle, each port is able to generate a transaction that will pass through a pipeline, a FIFO, and a back-pressured pipeline. If destined for a different port, the transaction will travel the ring to its destination, flow through another pipeline, arbitrate with local traffic, flow through another pipeline and into another FIFO.

A transaction is a collection of signals, or *fields*. The transaction flow defines the path by which these fields travel through the machine. Unlike structures, transactions "change shape" as their fields are operated upon by *transaction logic*. Each field travels along its flow from the point at which is is produced to the downstream locations in which it is consumed.

```
$ANY = $select ? /in1$ANY : /in2$ANY;
```

Fig. 2. Example flow expression for a multiplexer

Transaction flow is supported in TL-Verilog by one very simple mechanism. $ANY can be used in a logic expression in place of an output signal and one or more input signals. The expression must be one that transfers an $ANY input to the $ANY outputs, unaltered. An example would be the ternary operator, shown in Fig. 2, which implements a multiplexer, selecting one of two possible input transactions. We call such an expression a *flow expression*.

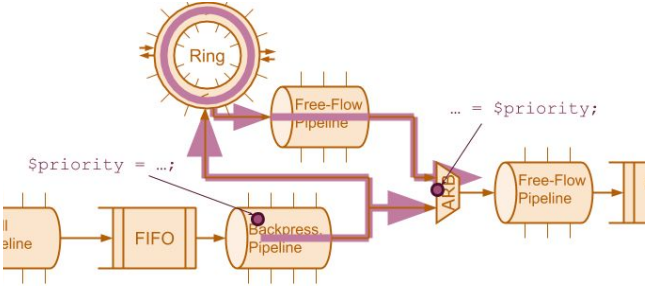

Fig. 3. Flow of a field

Flow expressions can be chained to form a directed graph that defines a flow like the one in Fig. 1. Having established such a flow, fields of the transaction can be assigned anywhere along the flow and consumed anywhere downstream in the flow. They will traverse the flow from their assignments to their downstream uses, as illustrated for the field $priority in Fig. 3.

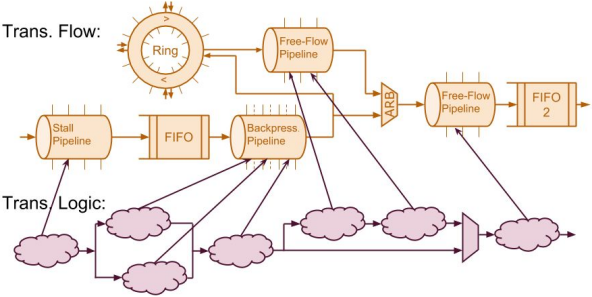

Fig. 4. Decoupling of flow logic and transaction logic

This creates a decoupling of *flow logic* and transaction logic, depicted in Fig. 4, similar to the decoupling of timing and behavior in timing-abstract design. Logic expressions can be moved upstream or downstream along their flows by cutting and pasting expressions. While this may affect the implementation dramatically, overall behavior remains unchanged. An expression, of course, cannot be moved upstream of its producers nor downstream of its consumers. The flow takes care of the sequential staging and steering of the fields, and the transaction logic is merely combinational.

### III. LEXICAL REENTRANCE

A very subtle, but very powerful property of TL-Verilog is something termed, "lexical reentrance." This refers to the ability to define content in a scope, change scope (in program order), then reenter the earlier scope to define more content. This simple property provides an important ability to separate concerns, similar in some respects to aspect-oriented programming.

Lexical reentrance is particularly valuable in the creation of libraries. Libraries, by their nature, provide generic functionality. They are instantiated in contexts that are less generic. Lexical reentrance enables the library components to provide generic aspects of functionality and the instantiating context to add more-detailed logic into the context established by the library components. The highly-flexible WARP-V CPU design [2], for example, takes advantage of lexical reentrance to provide CPU microarchitectural components that are independent of the instruction-set architecture (ISA) that is implemented.

We demonstrate in this work that lexical reentrance is particularly useful with transaction flows. It enables library components to define transaction flow, without defining transaction logic. The combinational logic of the transactions can be defined in the instantiating context. It is important to note that most bugs, particularly difficult bugs, tend to relate to sequencing and flow control. Combinational logic bugs are less common and are generally easier to verify and debug. As demonstrated in the next section, lexical reentrance enables complex, bug-prone flow logic to be defined in pre-verified library components.

### IV. FLOW LIBRARY AND ITS USE

While timing abstraction and transactions are well-defined and robustly supported by tools, the TL-Verilog language specification [3] does not define native features for modularity and reuse that would support the creation of a library. We utilized a proof-of-concept framework that uses the M4 macro preprocessor (and a bit of Perl). Our work is being used to aid in the specification of proper language features.

Our preliminary library consists of:

1. combinational flow components, like arbiters
2. pipelines with varying backpressure mechanisms, including phase-based backpressure [3]
3. FIFOs
4. routing components, like a ring, as well as a testbench macro for routing components

Flow control between library components is generally supported through a "ready/valid" interface, where the upstream component indicates "validity" of an available transaction, and the downstream component indicates its "readiness" to accept it. This backpressure mechanism controls the flow of transactions. For components providing this standard interface, there is no need for glue logic to stitch one interface to another. The components fit together like pieces of a jigsaw puzzle.

```
// Testbench
m4+router_testbench(/top, /ring_stop, |stall0, @1, |fifo2_out, @1,

/M4_RING_STOP_HIER
   m4+stall_pipeline(/ring_stop, |stall, 0, 3, /trans)
   m4+simple_bypass_fifo_v2(/ring_stop, |stall3, @1, |bp0, @1, 4,
   m4+bp_pipeline(/ring_stop, |bp, 0, 3, /trans)
   |bp3
      @1
         $local = /trans$dest != #ring_stop;
   m4+opportunistic_flow(/ring_stop, |bp3, @1, |bypass, @1, $local,
m4+simple_ring(/ring_stop, |ring_in, @1, |ring_out, @1, /top<>0$res

/ring_stop[*]
   m4+arb2(/ring_stop, |ring_out, @4, |bypass, @1, |arb_out, @1, /
   m4+simple_bypass_fifo_v2(/ring_stop, |arb_out, @1, |fifo2_out,
```

Fig. 5. Example transaction flow code

We implemented the flow from Fig. 1, together with its testbench, in just 13 lines of code, shown in Fig. 5. Eight of those lines are macro instantiations, four provide scope (replication, pipeline, and pipestage), and one is a logic expression. This model is sufficient for performance studies to explore issues such as the frequency with which arbitration forces local traffic onto the ring. It is also sufficient to enable the development of early verification models. (In fact, in this case, the model includes a testbench from our library.)

```
|calc
   @1
      $aa_sq[7:0] = $aa[3:0] ** 2;
      $bb_sq[7:0] = $bb[3:0] ** 2;
   @2
      $cc_sq[8:0] = $aa_sq + $bb_sq;
   @3
      $cc[4:0] = sqrt($cc_sq);
```

Fig. 6. Pipelined Pythagorean theorem calculation in TL-Verilog

To illustrate the extension from timing-abstract modeling to transaction-level modeling, we begin with an example used in [1] to illustrate timing abstraction. Fig. 6 implements Pythagoras' Theorem, computing $cc as the square root of $aa squared and $bb squared, distributed

over three pipeline stages. In this example, a transaction is a Pythagorean calculation. The logic expressions define the transaction logic, and the pipeline provides the transaction flow.

```
/ring_stop[*]
   |stall1
      @2
         /trans
            $aa_sq[7:0] = $aa[3:0] ** 2;
            $bb_sq[7:0] = $bb[3:0] ** 2;
   |bp1
      @1
         /trans
            $cc_sq[8:0] = $aa_sq + $bb_sq;
   |arb_out
      @1
         /trans
            $cc[4:0] = sqrt($cc_sq);
```

Fig. 7.   Pythagorean theorem calculation in flow

In the interest of consistency, we use this same (pre-verified) transaction logic as the transaction logic for our showcase example. Each transaction in our showcase will perform this computation. (More realistically, a router, like our showcase example might include transaction logic to decode packet headers.) The four logic statements could be added first into the beginning of the flow, enabling functional verification activities to verify the computations now performed by each transaction. Then the statements can be properly positioned along the flow to support the physical implementation. Fig. 7 shows the repositioned Pythagorean logic. The computation is split across the stall pipeline and the back-pressured pipeline in the sending port and the second free-flowing pipeline in the receiving port. Note that lexical reentrance is utilized to place this logic within the context established by our library macros.

## V. RESULTS

We quantify the impact of transaction-level design on our showcase example to illustrate the fundamental contributions of transaction-level design methodology.

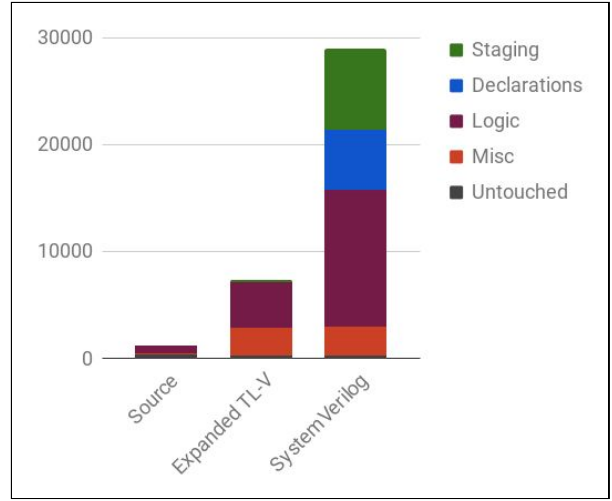

Fig. 8.   Characters of code from flow example (excluding comments and whitespace)

Fig. 8 provides code size data corresponding to the model in Fig. 5 (plus a module interface and include statements). The source code is expanded first by macro substitutions into pure TL-Verilog code and second by Redwood EDA's TL-Verilog compiler, SandPiper™, which generates the logic to stage and steer the signals through their flows.

Lexical reentrance and transaction-level design are responsible for some of our ability to efficiently utilize library components. Components that simply pass data through without any transaction logic, such as the ring, FIFOs, and arbiter could be implemented in SystemVerilog. In fact SystemVerilog components can be utilized in TL-Verilog with similar benefits to using TL-Verilog macros. The need to steer, stage, pack, and unpack fields at the inputs and outputs is eliminated. Components, like the various pipelines, on the other hand, are intended to contain transaction logic and cannot be implemented generically without lexical reentrance.

Transaction-level design, therefore, enables a substantial portion of the 6.8x code reduction from modularity, and it can reasonably be credited with the additional 4.0x reduction versus SystemVerilog. It should be noted that the SystemVerilog code is machine-generated, so it is not a perfect comparison versus a hand-coded SystemVerilog model, however the generated SystemVerilog conforms to an industry-recommended coding methodology with strong focus on readability.

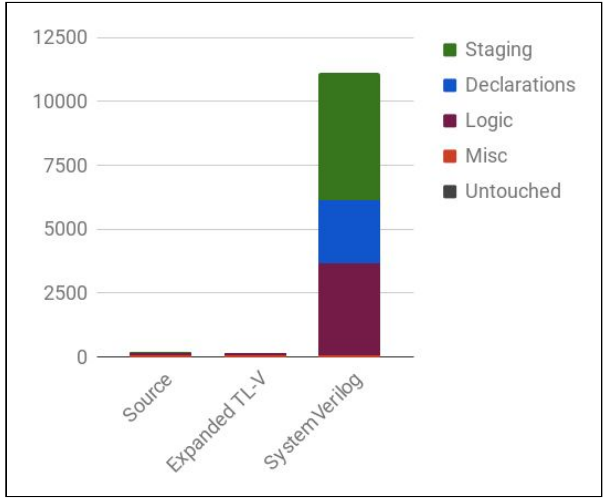

Fig. 9.  Characters of code from transaction logic

Fig. 9 characterizes the code resulting from the addition of the Pythagorean transaction logic. There is no macro expansion in this code. It generates 73x the SystemVerilog code, including signal declarations, staging logic, and steering logic (in flow expressions).

## VI. CONCLUSION

Our work demonstrates that rapid top-down design is feasible. We presented a library of generic design and verification components using TL-Verilog and M4 that enables quick microarchitecture and testbench construction. We illustrated that this early performance modeling and verification modeling can be extended with transaction logic to implement detailed RTL within the same fully-synthesizable modeling language and environment.